\documentclass[12pt]{iopart}

\usepackage{iopams}
\usepackage[normalem]{ulem}
\usepackage{graphicx}
\usepackage{array}

\def\lsim{\raise0.3ex\hbox{$\;<$\kern-0.75em\raise-1.1ex
\hbox{$\sim\;$}}}
\def\gsim{\raise0.3ex\hbox{$\;>$\kern-0.75em\raise-1.1ex
\hbox{$\sim\;$}}}

\begin{document}

{\hbox{YITP-SB-06-36}}
{\hbox{UAB-FT- 608}}
{\hbox{hep-ph/0609094}}

\vspace*{.25in}
\title{Probing long-range leptonic forces with solar
and reactor neutrinos}

\author{M C Gonzalez-Garcia$^{a,b}$, P C~de Holanda$^{c,d}$, E Mass\'o$^e$ and 
R~Zukanovich Funchal$^c$} 

\address{$^a$ IFIC, Universitat de Val\`encia - C.S.I.C., Apt 22085,
  E-46071 Val\`encia, Spain}
\address{$^b$ C.N.~Yang Institute for Theoretical Physics,\\
  SUNY at Stony Brook, Stony Brook, NY 11794-3840, USA} 
\address{$^c$
  Instituto de F\'{\i}sica, Universidade de S\~ao Paulo, C.\ P.\
  66.318, 05315-970 \\ S\~ao Paulo, Brazil}
\address{$^d$
  Instituto de F\'{\i}sica Gleb Wataghin, Universidade Estadual de
  Campinas, C.\ P.\ 6165, 13083-970 \\ Campinas, Brazil}
\address{$^e$ Departament de F\'{\i}sica and Institut de F\'{\i}sica 
  d'Altes Energies,\\
  Universitat Aut\`onoma de Barcelona, 08193 Bellaterra, Spain }

\eads{\mailto{concha@insti.physics.sunysb.edu}, \mailto{holanda@fma.if.usp.br},
  \mailto{zukanov@if.usp.br}}

\begin{abstract}
  In this work we study the phenomenological consequences of the
  existence of long-range forces coupled to lepton flavour numbers in
  solar neutrino oscillations.  We study {\sl electronic} forces
  mediated by scalar, vector or tensor neutral bosons and analyze
  their effect on the propagation of solar neutrinos as a function of
  the force strength and range.  Under the assumption of one mass
  scale dominance, we perform a global analysis of solar and KamLAND
  neutrino data which depends on the two {\it standard} oscillation
  parameters, $\Delta m^2_{21}$ and $\tan^2\theta_{12}$, the force
  coupling constant, its range and, for the case of scalar-mediated
  interactions, on the neutrino mass scale as well. We find that,
  generically, the inclusion of the new interaction does not lead to a
  very statistically significant improvement on the description of the
  data in the most favored MSW LMA (or LMA-I) region.  It does,
  however, substantially improve the fit in the high-$\Delta m^2$ LMA
  (or LMA-II) region which can be allowed for vector and scalar
  lepto-forces (in this last case if neutrinos are very hierarchical)
  at 2.5 $\sigma$.  Conversely, the analysis allows us to place
  stringent constraints on the strength versus range of the leptonic
  interaction.
\end{abstract}
\maketitle

\section{Introduction}
\label{sec:intro}

Many extensions of the standard model of particle physics predict the
existence of new forces; either generating deviations of the
gravitational law at short distances or predicting low mass particles
whose exchange will induce new forces at long distances, generally
violating the equivalence principle~\cite{damour,fischbach}.  A number
of experiments have been searching for such new forces but no evidence
has yet been found. These null results, nevertheless, provide very
significant constraints on particle physics models, gravitational
physics and even on cosmological speculations~\cite{fischbach,experiment}.

We know that there are two long-range forces in nature, namely the
electromagnetic and the gravitational force, but there is no a priori
reason why only these two should exist. Since the seminal work of Lee
and Yang \cite{lee-yang} it has become standard to consider long-range
forces coupling to baryon and/or lepton number. These would definitely
violate the universality of free fall and thus can be tested by
E\"otv\"os-type experiments, as pointed out in \cite{lee-yang}. In
particular, Okun \cite{okun} used this idea to establish a bound on
the strength of an hypothetical vectorial leptonic 
force; see \cite{dolgov} for a recent
review.  He obtained the bound on the ``fine structure'' constant
\begin{equation}
k_V< 10^{-49}\, .
\label{okun}
\end{equation}

The lepton flavour symmetries, as several solar, atmospheric,  
as well as KamLAND, K2K and MINOS  neutrino experiments indicate,
cannot be exact in nature. If an electronic force exists, we may
expect it to be of arbitrary but finite range. Notice that when the
range is less than the Earth-Sun distance, the bound (\ref{okun}) is 
no longer valid. Other experiments~\cite{fischbach,experiment} using
the Earth instead of the Sun as the electronic source derive bounds,
which, however, are much less strict than (\ref{okun}).

Recently, it has been shown that neutrino physics allow to place quite
stringent bounds on these new interactions.  The impact of vector-like
long-range forces coupled to lepton flavour numbers on solar and
atmospheric neutrino experiments has been investigated in \cite{gm2}
and \cite{mohanty}, respectively.  It has been estimated in \cite{gm2}
that the fine structure constant coupled to electronic number $L_e$ 
$k_V(e)<6.4 \times 10^{-54}$ to be compatible with solar neutrino
oscillations.  According to \cite{mohanty} atmospheric neutrino data
can constrain the strength of forces coupled to $L_e-L_\mu$, $k_V
(e\mu)\leq 5.5 \times 10^{-52}$, and to $L_e-L_\tau$, $k_V (e\tau)\leq
6.4 \times 10^{-52}$ (at 90\% CL) when the range of the force is the
Earth-Sun distance. Also, new forces coupling individually to the
muonic or tauonic number can be constrained, producing, however,
weaker bounds.  In particular, primordial nucleosynthesis
considerations provide the bound $k_V(\mu,\tau)<1.8 \times
10^{-11}$~\cite{gm1}.

In the present paper, we would like to continue the line of research
started in \cite{gm2}.  We perform a global analysis of the solar and
KamLAND neutrino data with the objective of constraining new
electronic forces. Since long-range forces can be mediated by the
exchange of massless or very light bosons of spin $J=0,1$ and 2, we
study all these cases. In the next Section the formalism to take into
account the effects of scalar (S), vector (V) and tensor (T) 
forces in neutrino 
oscillations is worked out. In Section \ref{sec:const} we perform 
the analysis and we devote the final Section to our conclusions.

\section{Formalism: Effects in Solar Neutrino Oscillations}
\label{sec:solarosc}

If there is a new force coupled to the leptonic flavour numbers, its
presence will affect neutrino oscillations when neutrinos travel
through regions where a flavour dependent density of leptons is
present as it is the case for solar neutrinos.
 
As we will see more in detail, how the effect of the new interaction
on the oscillation pattern depends on its Lorentz structure.
Nevertheless, it can be easily proved that for scalar, vector or
tensor interactions of large enough range the modification of the
evolution equation can always be casted in terms of a unique function
which solely depends on the background density of leptons -- the
source of the force -- and the range of the interaction.

In particular, in the Sun, we can define a {\sl universal} function
$W$ determining the effect of a new force coupled uniquely to $L_e$
at a point $r$ from the center of the Sun:
\begin{equation}
W(r) = \int_0^{R_{\rm Sun}} d^3 \rho \ n_e(\rho) \frac{e^{-\, |\vec \rho -
\vec r|/\lambda}}{|\vec \rho - \vec r|}\, ,
\label{W}
\end{equation}
where $R_{\rm Sun}$ is the radius of the Sun and $n_e(r)$ is the
electron number density in the medium. The range of the interaction is
$\lambda=1/m$, where $m$ is the exchanged boson mass.  We can easily
do the angular integration in (\ref{W}) and get
\begin{equation}
W(r) = \frac{2\pi\lambda}{r} \int_0^{R_{\rm Sun}} d\rho \ n_e(\rho) \ \rho
\left[ e^{-\, |\rho - r|/\lambda} \, - \, e^{-\, (\rho + r)/\lambda} \right]\, .
\label{W2}
\end{equation}

\begin{figure}[t]
\hglue 3.2cm
\includegraphics[width=4.5in]{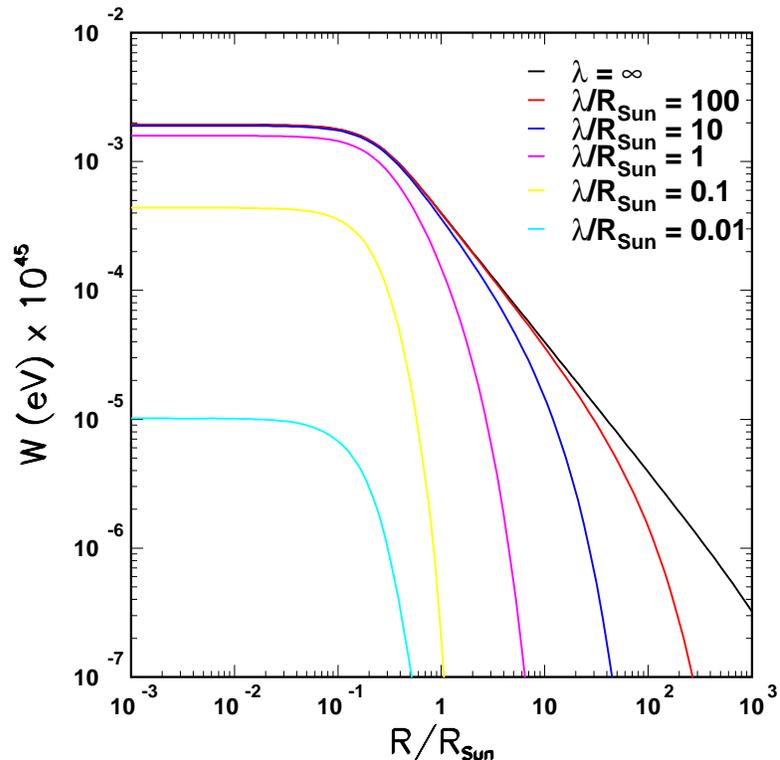}
\vglue -1cm
\caption{Leptonic {\sl potential} function $W(r)$ due to the density
  of electrons in the Sun as a function of the distance from the solar
  center in units of $R_{\rm Sun}$ for various ranges $\lambda$.}
\label{fig:pot}
\end{figure}

In Figure~\ref{fig:pot} we show the function $W(r)$ in the Sun as a
function of the distance in units of $R_{\rm Sun}$ for various ranges
$\lambda$. As seen in the figure, inside the Sun and up to 0.1 $R_{\rm
  Sun}$, $W$ is constant independently of the range of the force
considered. After that, it drops according to the force range and at
the position of the Earth, at about 200 $R_{\rm Sun}$ from the center
of the Sun, $W \lsim 2 \times 10^{-39}$ eV.

In order to see the effect of the new $L_e$-coupled force in the
evolution of solar neutrinos we have to take into account that at
present the minimum joint description of atmospheric~\cite{atm},
K2K~\cite{k2k}, MINOS~\cite{MINOS},
solar~\cite{chlorine,sagegno,gallex,sk,sno,sno05} and
reactor~\cite{kamland,chooz} data requires that all the three known
neutrinos take part in the oscillations.  The mixing parameters are
encoded in the $3\times 3$ lepton mixing matrix which can be
conveniently parametrized in the standard form
\begin{equation}
    U= \left(    \begin{array}{ccc}
    1&0&0 \\
    0& {c_{23}} & {s_{23}} \\
    0& -{s_{23}}& {c_{23}}  
    \end{array}\right)
\left(    \begin{array}{ccc}
    {c_{13}} & 0 & {s_{13}}e^{i {\delta}}\\
    0&1&0\\
    -{ s_{13}}e^{-i {\delta}} & 0  & {c_{13}}\\ 
    \end{array}\right)
\left(    \begin{array}{ccc}
    c_{21} & {s_{12}}&0\\
    -{s_{12}}& {c_{12}}&0\\
    0&0&1\cr
    \end{array}\right)\, ,
\end{equation}
where $c_{ij} \equiv \cos\theta_{ij}$ and $s_{ij} \equiv
\sin\theta_{ij}$. 

According to the current data, the neutrino mass squared 
differences can be chosen so that 
\begin{equation} \label{eq:deltahier}
    \Delta m^2_\odot = \Delta m^2_{21} \ll 
|\Delta m_{31}^2|\simeq|\Delta m_{32}^2|
=\Delta m^2_{\rm atm}.
\end{equation}
As a consequence of the fact that $\Delta m^2_{21}/\vert \Delta
m^2_{31}\vert \approx 0.03$, for solar and KamLAND neutrinos, the
oscillations with the atmospheric oscillation length are completely
averaged and the interpretation of these data in the neutrino
oscillation framework depends mostly on $\Delta m^2_{21}$,
$\theta_{12}$ and $\theta_{13}$ (while atmospheric and long baseline neutrino
oscillations are controlled by $\Delta m^2_{32}$, $\theta_{23}$ and
$\theta_{13}$).  Furthermore, the negative results from the CHOOZ
reactor experiment~\cite{chooz} imply that the mixing angle connecting
the solar and atmospheric oscillation channels, $\theta_{13}$, is
severely constrained ($\sin^2\theta_{13} \leq 0.041$ at 3$\sigma$
\cite{haga}).  Altogether, it is found that the 3-$\nu$ oscillations
effectively factorize into 2-$\nu$ oscillations of the two different
subsystems: solar (and reactor) and atmospheric (and long baseline). 

Because the  new interaction is flavour diagonal and its effect 
in the evolution of atmospheric neutrinos  does not modify the 
hierarchy (\ref{eq:deltahier}),  the 2$\nu$ oscillation  
factorization still holds. Thus, in general, one can write the solar neutrino 
evolution equation in the presence of the new $L_e$-coupled force as
\begin{equation} 
i \frac{d}{dr} 
\left(\begin{array}{c} \nu_e \\
\nu_\mu     \end{array}\right)=\left[
\frac{1}{2E_\nu}
     \mathbf{U}_{\theta_{12}}
 \; \mathbf{M^2} \; 
     \mathbf{U}_{\theta_{12}}^\dagger +
\left(    \begin{array}{cc}
        V(r) & ~0 \\
        \hphantom{-} 0 & ~0 
    \end{array}\right)\right]
\left(\begin{array}{c} \nu_e \\
\nu_\mu     \end{array}\right),
\label{eq:evol}
\end{equation}
where 
\begin{equation}
    \mathbf{U}_{\theta_{12}}=\left(    \begin{array}{cc}
    \cos \theta_{21} & {\sin \theta_{12}}\\
    -{\sin \theta_{12}}& {\cos \theta_{12}}
\end{array}\right),
\label{eq:mixing}
\end{equation}
and $\mathbf{M}$ and $V(r)$ will depend on the Lorentz structure 
of the new leptonic interaction as we describe next:

\subsection{Scalar  Interaction ($J=0$)}
If the new force is mediated by a neutral spin $J=0$  particle, $\phi$, 
the interaction Lagrangian for the neutrinos will take the form: 
\begin{equation}
\label{scalarL}
L = - g_0 \phi \bar \psi_\nu   \psi_\nu \, .
\end{equation}
We now use that the electron density can be taken to be the (static)
source of the $L_e$-coupled interaction and that the neutrino will
experience the force due to this source as long as within its range.
In this approximation the effect of the new interaction can be
included in the neutrino evolution equation as in (\ref{eq:evol})
with:
\begin{equation}
\fl
    \mathbf{M}=
\left(    \begin{array}{cc}
    m_1 & 0 \\
    0   & m_2
\end{array}\right)
- 
\mathbf{U}_{\theta_{12}}^\dagger\,
\left(  \begin{array}{cc}
    M_{\rm{S}}(r) & 0 \\
    0   & 0
\end{array}\right) \,    \mathbf{U}_{\theta_{12}}
\qquad {\rm{and}} \qquad V(r)=V_{\rm{CC}}(r),
\label{eq:scalar}
\end{equation}
where $m_{1,2}$ are the neutrino masses in vacuum,
$V_{\rm{CC}}(r)=\sqrt{2} \, G_F n_e(r)$ is the
Mikheyev-Smirnov-Wolfenstein (MSW) potential and $M_{\rm{S}}(r)=k_S(e)
W(r)$ is the new scalar-leptonic-force-induced mass term.  Here
$k_S(e)=\displaystyle\frac{g_0^2}{4\pi}$.  Different from the vector
and tensor cases, the $\nu_e$ survival probability will not only
depend on $\Delta m_{21}^2=m_2^2-m_1^2$ but also on the value of the
absolute neutrino mass scale $m_1$, which is still not experimentally
known \cite{Bilenky:2002aw}. Notice also that the term $M_S(r)$
has the same sign for neutrinos and antineutrinos.

\subsection{Vector  Interaction ($J=1$)}
If there is a new $L_e$-coupled vector force mediated by a neutral
vector boson $A_\alpha$, the interaction Lagrangian for the neutrinos
is:
\begin{equation}
\label{vectorL}
L = - g_1 A_\alpha \bar \psi_\nu  \gamma^\alpha \psi_\nu \; .
\end{equation}

In this case the neutrino evolution is described by (\ref{eq:evol}) with
\begin{equation}
    \mathbf{M}=\left(    \begin{array}{cc}
    m_1 & 0 \\
    0   & m_2
  \end{array}\right)
\qquad  {\rm and } \qquad V(r)=V_{\rm CC}(r)+V_{\rm V}(r),
\label{eq:vector}
\end{equation}
where $V_{\rm{V}}=k_V(e) W(r)$ is the lepto-vector potential with 
$k_V(e)=\displaystyle\frac{g_1^2}{4\pi}$. 
As a consequence of the vector structure of the force, the
new leptonic potential adds to the MSW potential with the same energy 
dependence and sign. It will accordingly flip sign when describing  
antineutrino oscillations.

\subsection{Tensor Interaction ($J=2$)}
For a $L_e$-coupled force mediated by a tensor field of spin $J=2$,
$\chi_{\alpha\beta}$, the interaction Lagrangian for the neutrinos is
given by
\begin{equation}
\label{tensorL}
L= - \frac{g_2}{2}\ \chi_{\alpha\beta} 
\left[ \bar \psi_\nu  \gamma^\alpha i  \partial^\beta \psi_\nu
-  i  \partial^\alpha  \bar \psi_\nu \gamma^\beta \psi_\nu
\right]\, ,
\end{equation}
where the coupling constant $g_2$ has dimensions  $1/E$. 

In this case the neutrino evolution is described by (\ref{eq:evol}) with
\begin{equation}
    \mathbf{M}=\left(    \begin{array}{cc}
    m_1 & 0 \\
    0   & m_2
 \end{array}\right)
\qquad  {\rm and} \qquad V(r)=V_{\rm CC }(r)- V_{\rm T }(r).
\label{eq:tensor}
\end{equation}
$V_{\rm{T}}(r)=E_\nu k_T(e) W(r)$ is the tensorial potential and $E_\nu$ 
is the neutrino energy. 
Here $k_T(e)=m_e\displaystyle\frac{g_2^2}{4\pi}$ and $m_e$ is 
the mass of the electron.

Finally, let us remark that, as for the case of a scalar leptonic force,
a tensor force is always symmetric when changing from neutrinos
to antineutrinos. This is obvious from (\ref{tensorL}) since what
we have coupled to the tensor field $\chi$ is in fact the energy
momentum tensor of the leptons, which has to be symmetric under the
exchange of particles and antiparticles.

\begin{figure}[hbt]
\hglue 3.2 cm
\includegraphics[width=4.5in]{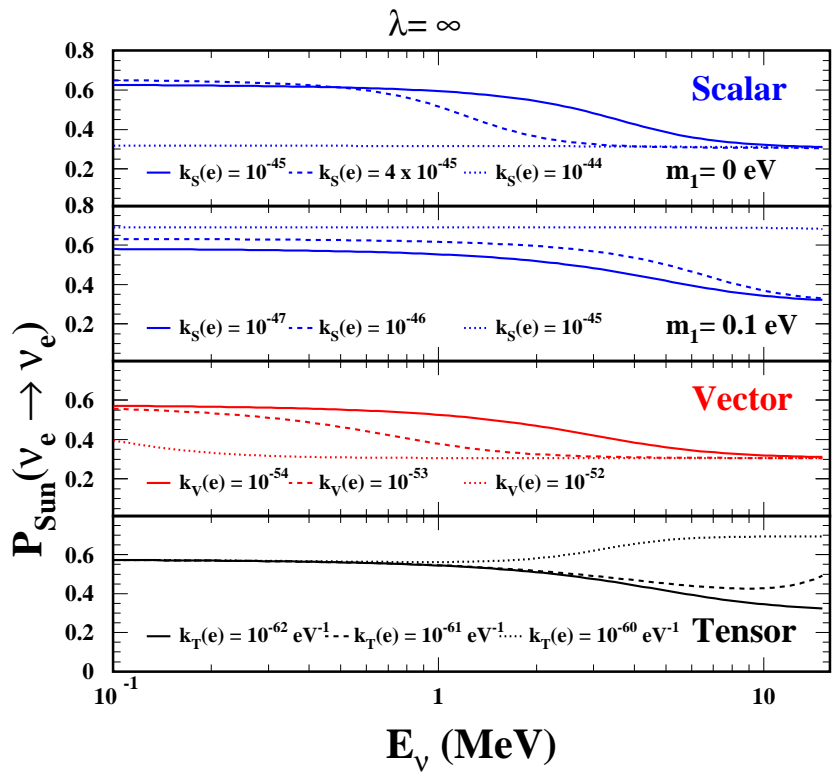}
\vglue -1.5cm
\caption{Survival probability of $\nu_e$ in the Sun as a 
function of the neutrino energy $E_\nu$ for an infinite range   
scalar (first two panels), vector (third panel) and 
tensor (lower panel) $L_e$-coupled force, for various 
values of the strength $k_i(e),\, i=S,V,T$. For all curves we have used 
$\tan^2\theta_{12}=0.44$ and $\Delta m^2_{21} =
7.9\times10^{-5}~{\rm{eV}}^2$.} 
\label{fig:infty}
\end{figure}

\begin{figure}[htb]
\hglue 1.2 cm
\includegraphics[width=4.5in]{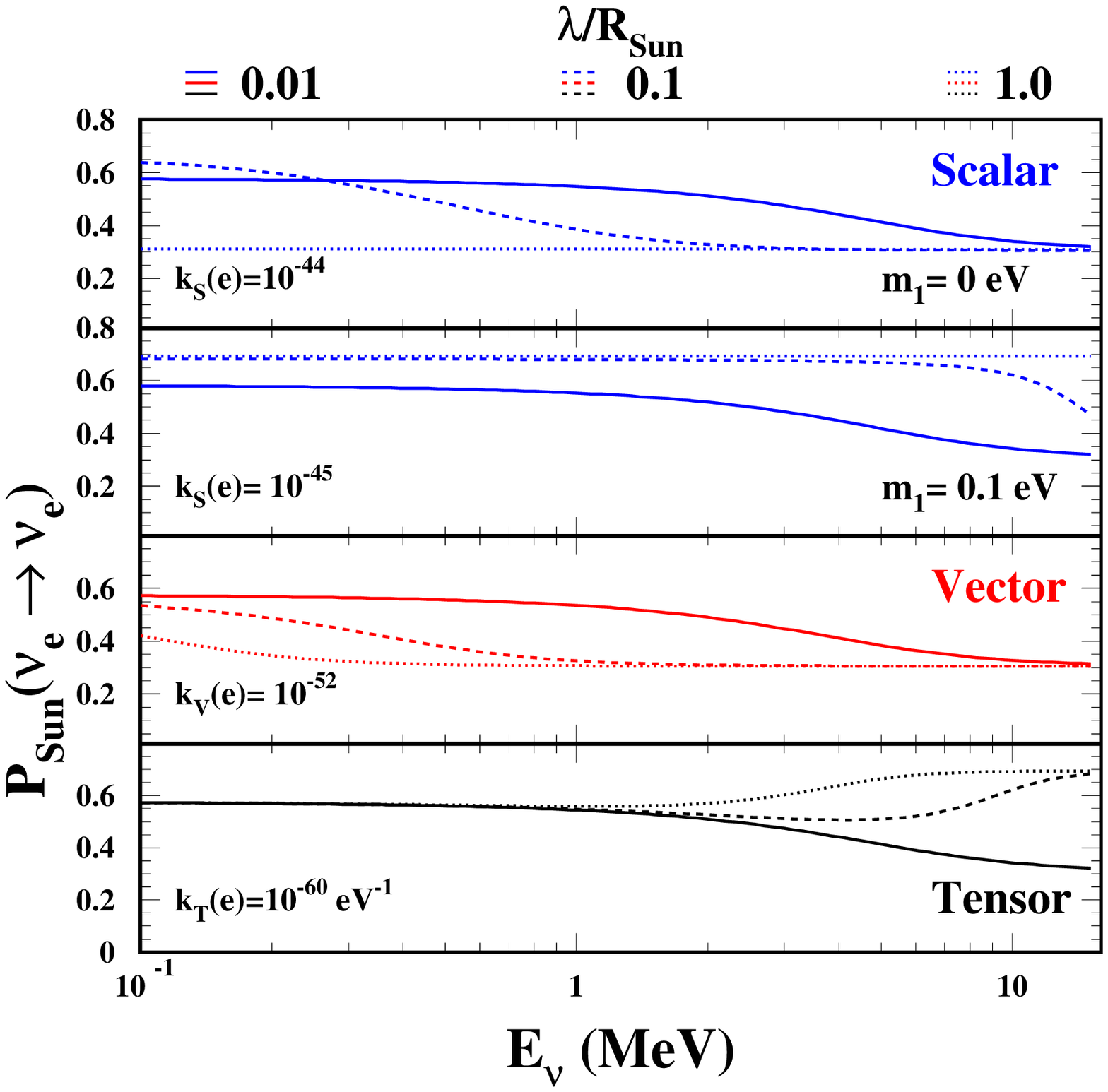}
\vglue -1.5cm
\caption{Same as Figure~\ref{fig:infty} but for 
$\lambda/R_{\rm Sun} = 0.01, 0.1$, and $1$.} 
\label{fig:finite}
\end{figure}

Before performing the data analysis we can have some idea of the size
of the effects of these long-range forces on neutrino oscillations 
by studying the value of the $\nu_e$ survival probability 
of solar neutrinos for different values of the coupling constant and range. 
The probability is obtained by solving (\ref{eq:evol}) along the
neutrino trajectory from its production point in the Sun to its detection
point on the Earth. We have verified that for the range of parameters of
interest the evolution in the Sun and from the Sun to the Earth is 
always adiabatic. It is interesting to remark that for the largest 
range forces the effective mixing angle at the surface of the Earth is 
still affected by the value of the leptonic potential produced by the Sun 
electron density. 

We show in Figure~\ref{fig:infty} the survival probability of solar 
$\nu_e$ at the sunny face of the Earth as a function of the neutrino 
energy $E_\nu$ for an infinite range force.  In the upper panels we show 
the scalar case for two extreme values of $m_1$, which correspond to 
strong mass hierarchy ($m_1=0$) and degenerate spectrum ($m_1=0.1$ eV), 
and some values of $k_S(e)$. From this figure we can foresee that for 
$m_1=0$, values of $k_S(e) \gtrsim 10^{-45}$--$10^{-44}$ will conflict 
with the existing solar neutrino data while for $m_1=0.1$ eV 
even smaller values of the coupling, $k_S(e) \gtrsim 10^{-46}$--$10^{-45}$,
will be ruled out. 
In the third panel we show the vector case for some values of $k_V(e)$. 
One expects from this that our analysis will lead to bounds $k_V(e) \lsim
10^{-54}$--$10^{-53}$.  Finally, in the lower panel the tensor
case is displayed.  In this case one expects that the data will
constrain $k_T(e) \lsim 10^{-61}-10^{-60}$ eV$^{-1}$.

The dependence of the $\nu_e$ survival probability on the range
$\lambda/R_{\rm Sun}$ can seen in Figure~\ref{fig:finite}. From this plot we
can see how the limit on $k$ will weaken as we allow for the force to
have a finite range.

\section{Constraints from Solar and Reactor Neutrino Data}
\label{sec:const}
We present in this section the results of the global analysis of
solar and KamLAND data for the long-range forces discussed in the
previous section. 

Details of our solar neutrino analysis have been described in previous
papers~\cite{oursolar,pedrosolar}.  We use the solar fluxes from
Bahcall {\it et al}~\cite{BS05}.  The solar neutrino data includes a
total of 119 data points: the Gallium~\cite{sagegno,gallex} (2 data
points) and
Chlorine~\cite{chlorine} (1 data point) radiochemical rates, the
Super-Kamiokande~\cite{sk} zenith spectrum (44 bins), and SNO data
reported for phase 1 and phase 2.  The SNO data used consists of the
total day-night spectrum measured in the pure D$_2$O (SNO-I) phase (34
data points)~\cite{sno}, plus the full data set corresponding to the
Salt Phase (SNO-II)~\cite{sno05}.  This last one includes the NC and
ES event rates during the day and during the night (4 data points),
and the CC day-night spectral data (34 data points). It is done by a
$\chi^2$ analysis using the experimental systematic and statistical
uncertainties and their correlations presented in~\cite{sno05},
together with the theoretical uncertainties. In combining with the
SNO-I data, only the theoretical uncertainties are assumed to be
correlated between the two phases.  The experimental systematics
errors are considered to be uncorrelated between both phases.

In the analysis of KamLAND, we neglect the effect of the long range
forces due to the small Earth-crust electron density in the evolution
of the reactor antineutrinos and we directly adapt the $\chi^2$ map as
given by the KamLAND collaboration for their unbinned rate+shape
analysis~\cite{kamhomepage} which uses 258 observed neutrino candidate
events and gives, for the standard oscillation analysis, a
$\chi^2_{\rm min}$=701.35. The corresponding Baker-Cousins $\chi^2$
for the 13 energy bin analysis is $\chi^2_{\rm min}=13.1/11$ dof.

\subsection{Scalar Interaction}

In the scalar case the analysis of solar and KamLAND data depends on 6
parameters: the two oscillation parameters, $\Delta m^2_{21}$ and
$\tan^2\theta_{12}$, the absolute neutrino mass scale $m_1$, the
coupling $k_S(e)$ and the range of the force $\lambda/R_{\rm Sun}$.
We have done our analysis fixing $m_1$ at two extreme values, $m_1=0$
and 0.1 eV.

For $m_1=0$ eV ($0.1$ eV) we find that the best fit point is
\begin{eqnarray}
\tan^2\theta_{12}=0.49\;  (0.44)   && \qquad  \Delta m^2_{21} = 
7.9\times10^{-5}~{\rm eV}^2 \; (7.9\times10^{-5}~{\rm eV}^2)     \nonumber \\  
\lambda/R_{\rm Sun}= 1.26\; ({\rm indet})   &&  \qquad 
k_S(e)=5\times 10^{-45}\; (0) \; \nonumber \\
 \Delta \chi^2_{\rm{min}}= -1.4\; (0)\, , &&
\label{eq:bestfitsca}
\end{eqnarray}
where the range of the interaction is indeterminate (indet) when
$k_{S}(e)=0$ and
$\Delta \chi^2_{\rm{min}}$ is the shift in 
$\chi^2$ compared to the best fit point in the absence of a new  
$L_e$-couple force for which 
\begin{eqnarray}
\tan^2\theta_{12}=0.44~ & \qquad {\rm and} \qquad & ~ \Delta m^2_{21} = 7.9\times10^{-5}~{\rm eV}^2 \, .
\label{eq:bestfitsm}
\end{eqnarray}
Thus we find that for highly hierarchical neutrinos the presence of
the long range force can lead to a small improvement of the quality of
the fit but this improvement is not statistically very significant
leading only to a decrease of less than two units in $\chi^2$ even at
the cost of introducing two new parameters. For degenerate neutrinos,
on the contrary, the presence of the long-range force does not improve
the agreement with the data.

\begin{figure}[t]
\vglue 1.2cm
\hglue 2.cm
\includegraphics[width=5.in]{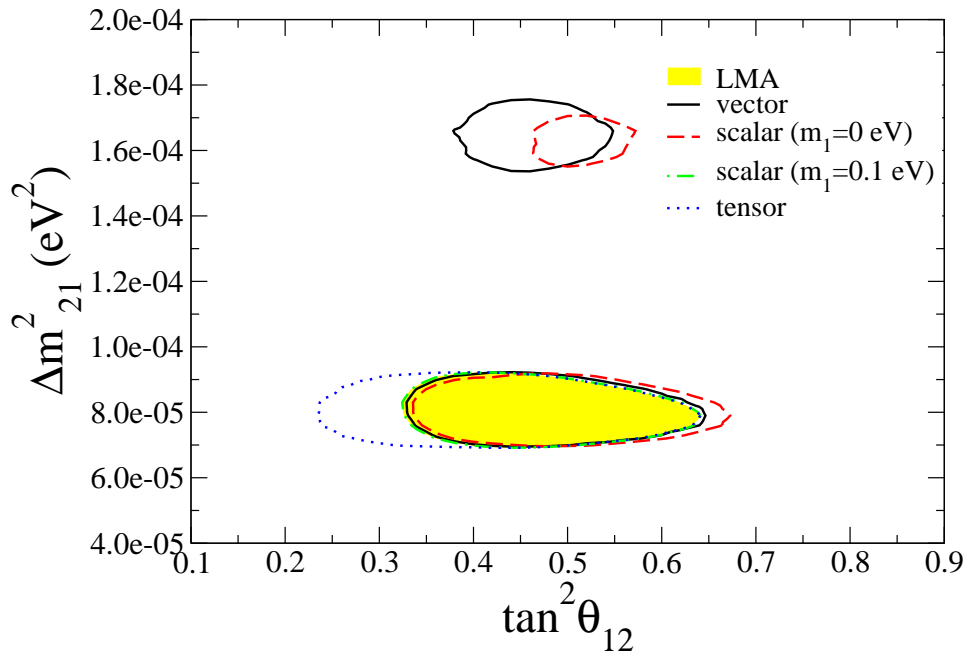}
\vglue 0.2cm
\caption{Allowed regions from the global analysis of  
solar plus KamLAND data in the $(\Delta m^2_{21},\tan^2\theta_{12})$ 
parameter space at 3$\sigma$ CL (2dof) for the different type
of long-range forces as labeled in the figure. 
The standard MSW allowed region (shaded area) is also shown for reference.}
\label{fig:oscregions}
\end{figure}

This is further illustrated in Figure~\ref{fig:oscregions} where we show 
the result of the global analysis of solar data plus KamLAND data in the 
form of the allowed two-dimensional regions at 3$\sigma$ CL in the 
$(\Delta m^2_{21}, \tan^2\theta_{12})$ plane after marginalization 
over $k_S(e)$ and $\lambda$. The standard MSW allowed region is also 
showed for reference.
As seen in the figure for any value of $m_1$ the presence of the 
long-range forces  has a mild (or null) impact on the 
allowed range  of $\Delta m^2_{21}$ and $\tan^2\theta_{12}$ in the 
low-$\Delta m^2$ island, known as large mixing angle region I (LMA-I).
Most interestingly, we  also find that in the presence of the long-range
scalar force the description of the solar data in the high-$\Delta m^2$ 
(LMA-II) region can be significantly improved if neutrinos 
are very hierarchical. As a consequence, 
in this case, there is a new allowed 
solution at the 98.9\% CL. The best fit point in this region 
is obtained for
\begin{eqnarray}
\tan^2\theta_{12}=0.5  \qquad && \qquad  \Delta m^2_{21} = 
1.6\times10^{-4}~{\rm eV}^2\nonumber \\  
\lambda/R_{\rm Sun}= 4.0\;  \qquad && \qquad 
k_S(e)=6.9\times 10^{-45}\; \; \nonumber \\
\Delta \chi^2_{\rm{min}}=  7.2\; &&
\label{eq:bestfitscahlma}
\end{eqnarray}
While this region is excluded at more than 4$\sigma$ for the standard MSW
oscillations it is allowed at 2.5$\sigma$ in the presence of a new scalar
force for a narrow band of the lepto-force parameters.  For example, 
for $\lambda=\infty$ the LMA-II region is allowed as long as $4.5\leq
(k_S(e)/10^{-45})\leq 8.0$~\footnote{The LMA-II region was found also
  for generic environmentally dependent solar neutrino masses
  \cite{ourmavas}}.

The improvement of the fit in the LMA-II region can be understood as
follows. For the range of scalar $L_e$-coupled lepto-forces that we
are considering the neutrino evolution in the Sun is always adiabatic.
Therefore the corresponding survival probability is determined by the
effective mixing angle at the neutrino production point, $r_0$, which
takes the form
\begin{equation}
\fl 
S2_{12,m}=
\frac{S2_{12} \Delta m^2_{21,S}}
{\sqrt{ [\Delta m^2_{21,S} C2_{12}-2 E_\nu V_{CC}(r_0)- M_S^2(r_0)
+ M_S(r_0) (m_1+m_2)]^2 +S2_{12}\Delta m^2_{21,S}}}  
\end{equation}
where we have defined $\Delta m^2_{21,S}=\Delta
m_{21}^2-M_S(r_0)\Delta m_{21}$ with $\Delta m_{21}=m_2-m_1$ and
$S2_{12}=\sin 2\theta_{12}$, and $C2_{12}=\cos 2 \theta_{12}$.  As
long as the term $M_S(r_0) (m_1+m_2)$ in the denominator is
subdominant, the net effect of $M_S(r_0)$ is to increase the effective
matter potential and correspondingly, one can achieve a good fit to
the solar data with higher values of $\Delta m^2_{21}$.  In this case,
the CL at which the LMA-II region is allowed is determined by KamLAND
data \cite{kamland} because the fit to the solar data cannot
discriminate between the LMA-I and LMA-II regions once scalar
$L_e$-coupled force effects are included.  Clearly this implies that
this solution will be further tested in the future by a more precise
determination of the antineutrino spectrum at KamLAND.

As $m_1$ increases the term $M_S(r_0) (m_1+m_2)$ becomes more important
and correspondingly  the quality of the fit within  the LMA-II 
region worsens. We find that for $m_1\gtrsim 0.02$ eV there is no 
LMA-II solution at 3$\sigma$.

\begin{figure}[t]
\hglue 2.1cm
\includegraphics[width=5.5in]{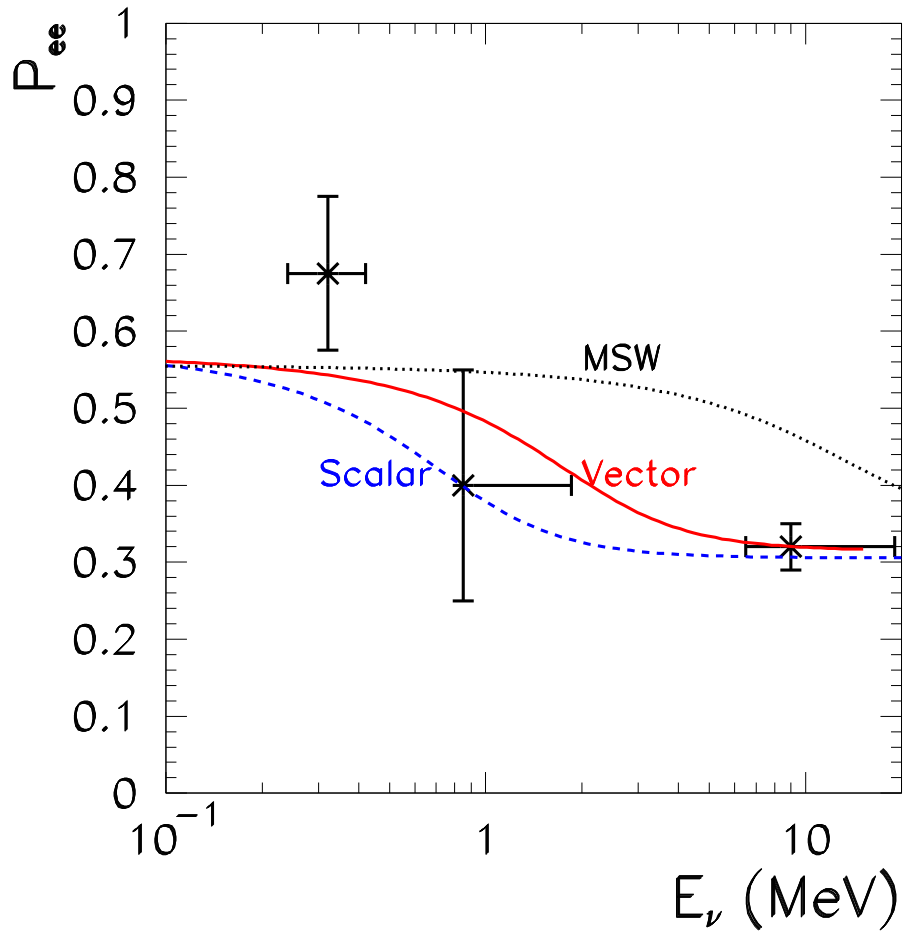}
\caption{$\nu_e$ survival probability in the Sun versus neutrino
  energy for the best fit point in the high-$\Delta m^2$ region in the
  presence of scalar lepto-force with parameters given in
  (\ref{eq:bestfitscahlma}) (full line) and for the vector lepto-force
  with parameters in (\ref{eq:bestfitvhlma}).  The dotted line is the
  survival probability for conventional MSW oscillations ($k_i=0$) for
  the same values of $\Delta m^2_{21}$ and $\theta_{12}$.  The data
  points are the extracted average survival probabilities for the low
  energy ($pp$), intermediate energy ($^7$Be, $pep$ and CNO) and high
  energy solar neutrinos ($^8$B and $hep$) from~\cite{Barger}.}
\label{fig:probhigh}
\end{figure}

We show in Figure~\ref{fig:probhigh} the survival probability for this
best fit point in the high-$\Delta m^2$ (LMA-II) region in the
presence of the scalar lepto-force together with the extracted average
survival probabilities for the low energy ($pp$) , intermediate energy
($^7$Be, $pep$ and CNO) and high energy solar neutrinos ($^8$B and
$hep$) from \cite{Barger}. For comparison we also show the survival
probability for conventional MSW oscillations for the same values of
$\Delta m^2_{21}$ and $\theta_{12}$.  From the figure it is clear that
the inclusion of the lepto-force, leads to an improvement on the
description of the solar data for all the energies being this more
significant for intermediate- and high-energy neutrinos.

\begin{figure}[htb]
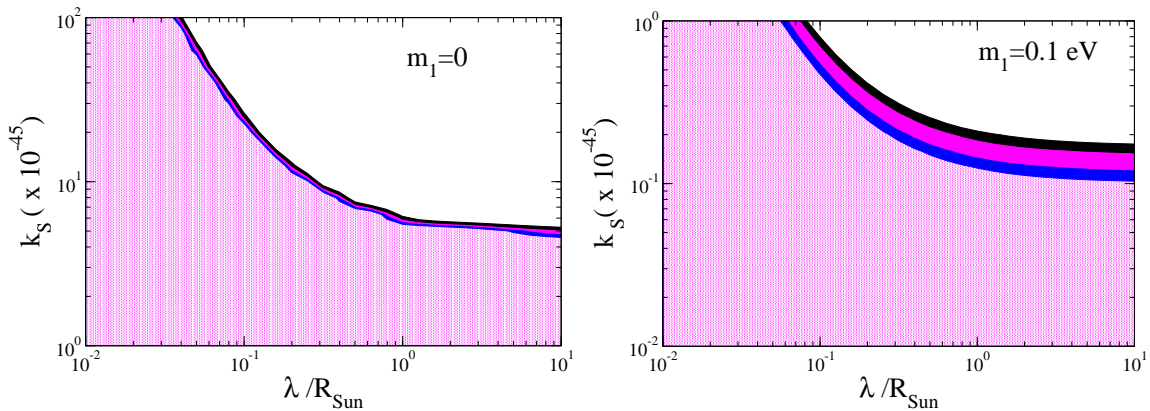

\vglue 1.cm
\hglue 1.1cm
\includegraphics[width=0.48\textwidth]{chi2l1l2.scalar1.eps}
\includegraphics[width=0.48\textwidth]{chi2l1l2.scalar2.eps}
\vglue 0.3cm
\caption{Allowed regions from the global analysis of solar plus
  KamLAND data in the $k_S(e)$ versus $\lambda/R_{\rm Sun}$ plane at
  90\%, 95\%, 99\% and 3$\sigma$ CL (2dof) for a $L_e$-coupled scalar
  force, with $m_1=0$ eV (left) and $m_1=0.1$ eV (right).}
\label{fig:krscalar}
\end{figure}

Conversely, the global analysis of solar and KamLAND data results into
constraints on the possible size of the new contribution to the
neutrino mass.  This is illustrated in Figure~\ref{fig:krscalar} where
we show the result of the global analysis in the form of the allowed
two-dimensional regions in the $(k_S(e),\lambda/R_{\rm Sun})$
parameter space after marginalization over $\Delta m^2_{21},
\tan^2\theta_{12}$ for the two reference values of $m_1$.

As seen in the figure, already at 90\% CL (or lower)  
the regions are connected to the standard $k_S(e)=0,\; \lambda=\infty$  
case. In other words, the analysis shows no evidence of any
long-range $L_e$-coupled scalar force contributing to the solar neutrino 
evolution. However, as expected,  there is a strong  correlation between 
the implied bound on the strength and the range of the 
interaction. As discussed in the previous section the 
tightest bound on the strength of the new interaction corresponds
to $\lambda=\infty$ (which effectively applies for 
$\lambda\gtrsim 10\, R_{\rm Sun}$)
\begin{eqnarray}  
&& k_S(e)\leq 5.0 \times 10^{-45} \quad \quad (g_0\leq 2.5\times 10^{-22})
\; \; \; \; \; \; \; {\rm for \; m_1=0 \; \rm eV}\, ,\\
&& k_S(e)\leq 1.5 \times 10^{-46} \quad \quad (g_0\leq 4.3\times 10^{-23})
\; \; \; \; \; \; \; {\rm for \; m_1=0.1 \; \rm eV}\, ,
\end{eqnarray}
at 3$\sigma$ (1dof).

\subsection{Vector Interaction}

If the new leptonic interaction is mediated by a vector boson 
the analysis of solar and KamLAND data depends on 
4 parameters: the two  oscillation parameters,  
$\Delta m^2_{21}$ and $\tan^2\theta_{12}$, the coupling $k_V(e)$ 
and the range of the force $\lambda/R_{\rm Sun}$.

In this case the best fit point corresponds to 
\begin{eqnarray}
\tan^2\theta_{12}=0.46\; \quad  && \quad \quad \Delta m^2_{21} = 
7.9\times10^{-5}~{\rm eV}^2     \nonumber \\  
\lambda/ R_{\rm Sun}= 2.5\times 10^{-3}\;  \quad && \quad \quad
k_V(e)=7.9\times 10^{-51}\;  \; \nonumber \\
\Delta \chi^2_{\rm{min}}= -0.8\; , &&
\label{eq:bestfitv}
\end{eqnarray}
which as for the scalar case represent a very small variation with
respect to the standard LMA solution as seen in
Figure~\ref{fig:oscregions}.

From Figure~\ref{fig:oscregions} we also find that the presence of the vector 
lepto-force can  improve the quality of the fit in the LMA-II region.
This is possible because the lepto-force also adds up to the MSW potential
and consequently the vacuum-matter transition condition 
$\Delta m^2_{21} \cos 2\theta = 2 E_\nu V(r_0)$, which has to occur
around $E_\nu\sim 1$ MeV to fit the data, can be achieved for a larger
value of $\Delta m^2_{21}$. 

So also in the vector case, there is a new allowed solution at 
2.5 $\sigma$ CL. The best fit point in this region is obtained for
\begin{eqnarray}
\tan^2\theta_{12}=0.46 \quad  &&  \quad \quad \Delta m^2_{21} = 
1.6\times10^{-4}~{\rm eV}^2\nonumber \\  
\lambda/R_{\rm Sun}= 4.0\; \quad  && \quad \quad 
k_V(e)=6.9\times 10^{-45}\; \; \nonumber \\
\Delta \chi^2_{\rm{min}}=  7.9\; &&
\label{eq:bestfitvhlma}
\end{eqnarray}  
The corresponding survival probability for this point is shown in
Figure~\ref{fig:probhigh}.  The LMA-II region is allowed for a narrow
band of $(k_V(e),\lambda/R_{\rm Sun})$ values. For example for
$\lambda=\infty$ the LMA-II region is allowed as long as $2.0\leq
(k_V(e)/10^{-54})\leq 30$.

As for the scalar lepto-force with hierarchical neutrinos, the CL at
which the LMA-II region is allowed is determined by KamLAND
data~\cite{kamland} and consequently this solution will be further
tested by this experiment.

\begin{figure}[htb]
\vglue 1.cm
\hglue 3.8cm
\includegraphics[width=.6\textwidth]{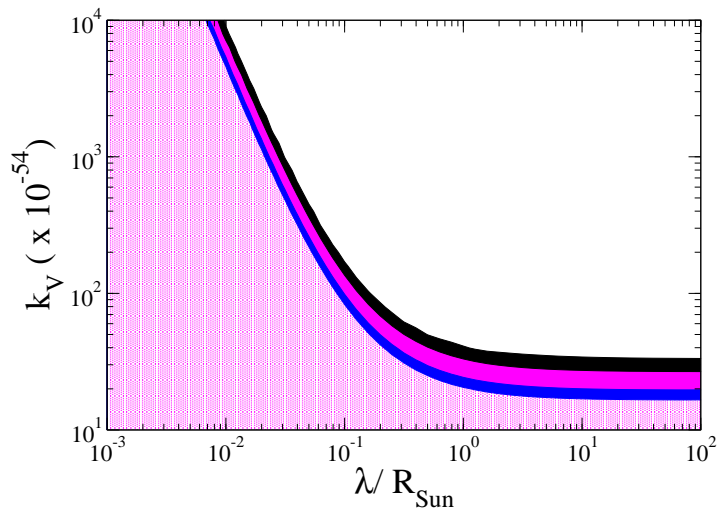}
\vglue 0.3cm
\caption{Allowed regions from the global analysis of  
solar plus KamLAND data in the $k_V(e)$ versus $\lambda/R_{\rm Sun}$ 
plane  at 90\%, 95\%, 99\% and 3$\sigma$ CL (2dof)  for 
a $L_e$-coupled vector force.} 
\label{fig:krvector}
\end{figure}

Concerning the allowed ranges of the lepto-force parameters, we show
in Figure~\ref{fig:krvector} the result of the global analysis in the
form of the allowed two-dimensional regions in the
$(k_V(e),\lambda/R_{\rm Sun})$ parameter space after marginalization
over $\Delta m^2_{21}, \tan^2\theta_{12}$. In particular we find that
for $\lambda=\infty$
\begin{equation}  
k_V(e)\leq 2.5 \times 10^{-53} \;\; (g_1\leq 1.8\times 10^{-26})\, ,
\end{equation}
at 3$\sigma$ (1dof).

\subsection{$J=2$ Interaction }

In the tensor case the analysis of solar and KamLAND data depends on 
4 parameters: the two oscillation parameters,  
$\Delta m^2_{21}$ and $\tan^2\theta_{12}$, the coupling $k_T(e)$ 
and the range of the force $\lambda/R_{\rm Sun}$.

We find the best fit point at 
\begin{eqnarray}
\tan^2\theta_{12}=0.44 \quad && \quad \quad \Delta m^2_{21}= 
7.9\times10^{-5}~{\rm eV}^2     \nonumber \\  
\lambda/R_{\rm Sun}=({\rm  indet}) \quad && \quad \quad
k_T(e)=0  \; \nonumber \\
\Delta\chi^2_{\rm{min}}=0 \;. &&
\label{eq:bestfitt}
\end{eqnarray}
Thus for the case of the $J=2$ interaction there is no improvement
on the quality of the fit compared to the standard MSW scenario.

From Figure~\ref{fig:oscregions} we see that the only effect of
the inclusion of this interaction is the enlargement of the 
LMA-I region for smaller values of the mixing angle. 
The inclusion of the $J=2$ lepto-force decreases
the effective matter potential (see (\ref{eq:tensor})) 
at higher energies. As a consequence, as seen in 
Figures~\ref{fig:infty} and~\ref{fig:finite}  the corresponding
survival probability grows for the high end of the 
solar neutrino spectrum as compared to the standard MSW
case. This behaviour can be compensated by a decrease of
the vacuum mixing angle and leads to the enlargement of
the allowed LMA-I region.   
However, because the effect of the $J=2$ lepto-force is to decrease 
the matter potential,  no LMA-II solution is found in this case. 

\begin{figure}[htb]
\vglue 1.cm
\hglue 3.8cm
\includegraphics[width=.6\textwidth]{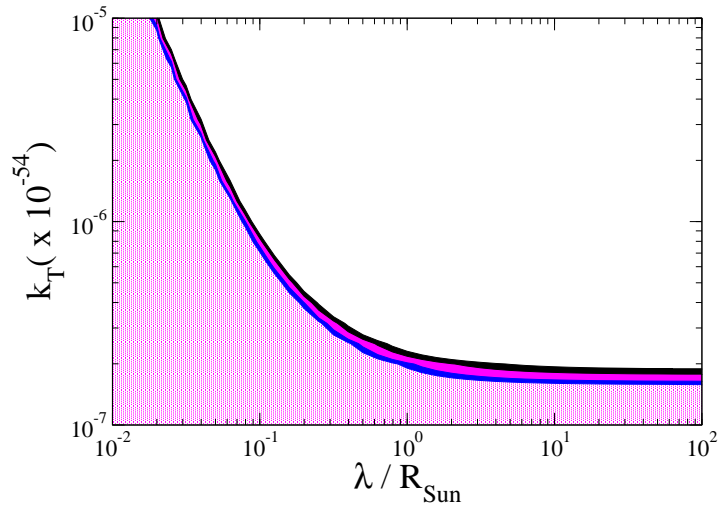}
\vglue 0.3cm
\caption{Allowed regions from the global analysis of  
solar plus KamLAND data in the $k_T(e)$ versus $\lambda/R_{\rm Sun}$ 
plane at 90\%, 95\%, 99\% and 3$\sigma$ CL (2dof)  for 
a $L_e$-coupled tensor force.} 
\label{fig:krtensor}
\end{figure}

Finally, in Figure~\ref{fig:krtensor}, we show the result of the global
analysis in the form of the allowed two-dimensional regions in the
$(k_T(e),\lambda/R_{\rm Sun})$  parameter space after marginalization over
$\Delta m^2_{21}$ and  $\tan^2\theta_{12}$. For  $\lambda=\infty$ we 
obtain the bound
\begin{equation}  
k_T(e)\leq 1.7 \times 10^{-60}\; {\rm eV}^{-1} 
\;\; (g_2\leq 6.5\times 10^{-33}\; {\rm eV}^{-1})\, , 
\end{equation}
at 3$\sigma$ (1dof).

\section{Conclusion and discussion}
\label{sec:disc}
In this paper be have investigated the effects of long-range
$L_e$-coupled forces in the oscillation of neutrinos in the Sun. Their
impact on neutrinos depend on the spin of the exchanged particle
leading to the new force. We have treated, in turn, the consequences
of scalar, vector and tensor forces.

In our study we have used data from solar neutrino experiments and
KamLAND. In the global fit, apart from the mass difference $\Delta
m^2_{21}$ and the mixing angle $\theta_{12}$, which are the usual
standard parameters, we have the new physics parameters $k_i$, the
strength of the interaction with $i$=S,V,T, and the range $\lambda$ of
the interaction. For the scalar case, we also need to assume the
absolute mass scale for neutrinos.

The result of the global fit is that there is not a significant
improvement in the description of the data in the preferred (in the
standard case) MSW LMA region (LMA-I). However, the addition of new
physics permits to ``recover" the high-$\Delta m^2$ region (LMA-II)
which is now allowed in the case of scalar and vector forces. Why this
is so only for these forces and not for the tensorial one can be
understood quite simply, as we have explained in the paper.

We have deduced limits on the strength of the forces for different
values of the range $\lambda$ (see Figures \ref{fig:krscalar},
\ref{fig:krvector} and \ref{fig:krtensor}). For an
infinite range ($\lambda=\infty$) $L_e$-coupled force, we get 
the following limits in terms of the ``fine
structure constants" $k_i$

\begin{eqnarray}
k_S(e) & \leq & 5.0 \times 10^{-45} \quad \quad (m_1=0 \; {\rm eV})\, ,
\label{scalar_conc1}\\ 
k_S(e) & \leq & 1.5 \times 10^{-46} \quad \quad (m_1=0.1 \; {\rm eV})\, ,
\label{scalar_conc2}\\ 
k_V(e) & \leq & 2.5\times 10^{-53}\, , \label{vector_conc}\\ 
k_T(e) & \leq & 1.7 \times 10^{-60}\; {\rm eV}^{-1}\, , \label{tensor_conc}
\end{eqnarray}
at 3$\sigma$ (1dof).

These bounds are practically the same for any $\lambda\gtrsim 10\,
R_{\rm Sun}$. For $\lambda\lesssim 10\, R_{\rm Sun}$, the bound slowly
worsens, and for $\lambda\sim 0.1\, R_{\rm Sun}$ we have that the bound on
$k_i$ is a factor less than 10 weaker than
(\ref{scalar_conc1})-(\ref{tensor_conc}). In   \cite{gm2} the
vectorial case was evaluated, not with the statistical rigor we have
proceeded in the present paper. A bound on $k_V(e)$ one order of
magnitude more stringent was found. Apart from extending the results
to include scalar and tensor forces, our bound (\ref{vector_conc})
supersedes the one in \cite{gm2}.

In deriving (\ref{scalar_conc1}) to (\ref{tensor_conc}) we have
assumed that the new force only couples to $L_e$ but the analysis can
be easily extended to forces coupled to any linear combination of the
lepton flavour numbers $L_e+\alpha L_\mu + \beta L_\tau$. For vector
and tensor forces it can be easily proved that as long as the
2$\nu$-oscillation factorization holds, the bounds will be the ones
given in (\ref{vector_conc}) and (\ref{tensor_conc}) but they
will constrain the combination $k_{V,T}=k_{V,T}(e)(1-\alpha
\cos2\theta_{23}- \beta \sin2\theta_{23})$.  In particular this means
that for the anomaly-free combinations $L_e-L_\mu$, and $L_e-L_\tau$
the bounds on the corresponding $k_{V,T}(e\mu)$ and $k_{V,T}(e\tau)$
become tighter by a factor $1+\cos2\theta_{23}\simeq 2$ and
$1+\sin2\theta_{23}\simeq 2$ respectively.  For the case of the scalar
interactions coupled to a general linear combination of lepton flavour
numbers, the combination of couplings bounded depends on $m_1$ and it
does not have such a simple expression as for the tensor and vector
cases.  But generically, provided that there is not a fine-tuned cancellation
between the masses and the new interaction terms, the bound will be of
the same order of magnitude as given in (\ref{scalar_conc1}) and
(\ref{scalar_conc2}).

A final issue we should care about is the possibility of screening of
leptonic charges by electronic neutrinos and antineutrinos from the
cosmological relict background.  As is explicitly demonstrated in
\cite{gm2}, screening plays no role at the relevant scales on the
order of the solar radius.


\ack
We are grateful to J.\ Garriga and J.\ A.\ Grifols for useful
discussions. This work is supported by National Science Foundation
grant PHY-0354776, by Spanish grants FPA-2004-00996 and FPA2005-05904,
by Catalan DURSI grant 2005SGR00916, by Funda\c{c}\~ao de Amparo \`a
Pesquisa do Estado de S\~ao Paulo (FAPESP) and by Conselho Nacional de
Ci\^encia e Tecnologia (CNPq).


\end{document}